\documentclass[10pt,final,twocolumn]{IEEEtran}
\usepackage{caption}
\usepackage[caption=false]{subfig}
\usepackage{amsmath}
\usepackage{amssymb}
\usepackage{amsthm}
\usepackage{mathtools}
\usepackage{graphicx}
\usepackage{float}
\usepackage{breqn}
\usepackage{bm}
\usepackage{url}
\usepackage{cite}
\usepackage{placeins}

\begin{document}

\newcommand\numberthis{\addtocounter{equation}{1}\tag{\theequation}}
\newcommand*{\TitleFont}{\usefont{\encodingdefault}{\rmdefault}{b}{n}\fontsize{12}{20}\selectfont}
\numberwithin{equation}{section}

\title{Spectral-Domain Computation of Fields Radiated by Sources in Non-Birefringent Anisotropic Media}
\author{Kamalesh~Sainath and Fernando L. Teixeira\thanks{The authors are with the ElectroScience Laboratory (ESL), Department of Electrical and Computer Engineering, The Ohio State University (OSU), Columbus, Ohio 43212 USA, (email: $\{$sainath.1@,teixeira@ece.$\}$osu.edu).}
\thanks{This work was supported by the NASA-NSTRF program and by the OSC.}}
\maketitle
\begin{abstract}
We derive the key expressions to robustly address the eigenfunction expansion-based analysis of electromagnetic (EM) fields produced by current sources within planar non-birefringent anisotropic medium (NBAM) layers. In NBAM, the highly symmetric permeability and permittivity tensors can induce directionally-dependent, but polarization independent, propagation properties supporting ``degenerate" characteristic polarizations, i.e. four linearly-independent eigenvectors associated with only two (rather than four) unique, non-defective eigenvalues. We first explain problems that can arise when the source(s) specifically reside within NBAM planar layers when using canonical field expressions. To remedy these problems, we exhibit alternative spectral-domain field expressions, immune to such problems, that form the foundation for a robust eigenfunction expansion-based analysis of time-harmonic EM radiation and scattering within such type of planar-layered media. Numerical results demonstrate the high accuracy and stability achievable using this algorithm.    
\end{abstract}

\section{Introduction}	
Environments with (locally) planar-layered profiles are encountered in diverse applications such as geophysical exploration, ground penetrating radar, conformal antenna design, and so on~\cite{sainath,sainath3,lambot1}. To facilitate electromagnetic (EM) radiation analysis in such environments, eigenfunction (plane wave) expansions (PWE) have long been used because of their relative computational efficiency versus brute-force numerical methods such as finite difference and finite element methods. Moreover, PWE can accommodate linear, but otherwise arbitrary anisotropic layers characterized by arbitrary (diagonalizable) 3$\times$3 material tensors~\cite{sainath}. This proves useful when rigorously modeling planar media simultaneously exhibiting both electrical and magnetic anisotropy, such as (i) isoimpedance beam-shifting devices and (to facilitate proximal antenna placement) ground-plane-coating slabs systematically designed via transformation optics (T.O.) techniques~\cite{sainath3,pendry5}, (ii) more practically realizable (albeit not necessarily isoimpedance) approximations to T.O.-inspired media such as metamaterial-based thin, wide-angle, and polarization-robust absorbers to facilitate (for example) radar cross section control~\cite{huangfu}, as well as (iii) numerous other media such as certain types of liquid crystals, elastic media subject to small deformations, and superconductors at high temperatures~\cite{boulanger1}. These named, amongst other, modeling scenarios share in common the potential presence of a particular class of anisotropic media in which the magnetic permeability ($\boldsymbol{\bar{\mu}}_r$) and electric permittivity ($\boldsymbol{\bar{\epsilon}}_r$) tensor properties are ``matched" to each other and hence together define media supporting four ``degenerate" plane wave eigenfunctions that, while possessing four linearly independent field polarization states (eigenvectors) as usual, share only two unique (albeit, critically still, \emph{non-defective}) eigenvalues~\cite{felsen}. Alternatively stated, propagation characteristics within such media are still (in general) dependent on propagation direction but \emph{independent} of polarization, eliminating ``double refraction" (``birefringence") effects~\cite{felsen,boulanger1}. Hence our proposed moniker ``Non-Birefringent Anisotropic Medium" (NBAM), rather than the ``pseudo-isotropic" moniker~\cite{boulanger1}. 

From an analytical standpoint, said PWE constitute spectral integrals exactly quantifying the radiated fields~\cite{sainath3}. Except for some very simple cases however, these expansions must almost always be evaluated by means of numerical quadratures or cubatures, whose robust computation (with respect to varying source and layer properties) is far from trivial and requires careful choice of appropriate quadrature rules, complex-plane integration contours, etc. to mitigate discretization and truncation errors as well as accelerate convergence~\cite{sainath,sainath3,mich2}. In addition to such considerations of primarily {\it numerical} character, a distinct problem occurs, due to said eigenvalue degeneracy, when sources radiate within NBAM layers. Indeed, this case requires proper analytical ``pre-treatment" of the fundamental spectral-domain field expressions to avoid two sources of ``breakdown": (i) Numerically unstable calculations (namely, divisions by zero) during the computation chain, as well as (ii) Corruption of the correct form of the eigenfunctions, viz. $z$exp[$ik_zz$] instead of the proper form exp[$ik_zz$], the former resulting from a naive, ``blanket" application of Cauchy's integral theorem to the canonical field expressions~\cite{hanson1,Weber1980}.

To this end, we first show the key results detailing the degenerate ``direct" (i.e., homogeneous medium) radiated fields in the ``principal material basis" (PMB) representation with respect to which the material tensors are assumed simultaneously diagonalized by an orthogonal basis~\cite{pendry5}.\footnote{\label{ortho}Note: The material tensor eigenvectors $\{\bold{\hat{v}}_1,\bold{\hat{v}}_2,\bold{\hat{v}}_3\}$ are \emph{not} to be confused with the field polarization eigenvectors.} Subsequently, we transform these PMB expressions to the Cartesian basis (the PWE's employed basis). Finally, we employ a robust, numerically-stable NBAM polarization decomposition scheme to obtain the Cartesian-basis direct field polarization amplitudes. The two-dimensional (2-D) Fourier integral-based PWE algorithm, resulting from implanting these derived field expressions into an otherwise highly robust PWE algorithm~\cite{sainath3}, comprises this paper's central contribution.

%Note that the derivation concerning the \emph{direct} fields (our interest) does not require our assuming a planar-layered medium; hence in %Section 3 and beyond we will assume the source resides within a homogeneous NBAM of material properties %$\{\boldsymbol{\bar{\epsilon}}_{r},\boldsymbol{\bar{\mu}}_{r} \}$. Since derivation of the key results is motivated by directly incorporating %the resultant expressions into spectral-domain planar-layered algorithms however (say, media stratified along $z$), we will assume fixed %transverse wave numbers $\{k_x,k_y\}$ at which we examine the spectral-domain direct fields. 

\section{\label{intro2}Problem Statement}

We assume the $\exp{-i\omega t}$ convention in what follows.  
Within a homogeneous medium of material properties $\{\boldsymbol{\bar{\epsilon}}_{r},\boldsymbol{\bar{\mu}}_{r} \}$, the electric field $\bm{\mathcal{E}}(\bold{r})$ radiated by electric ($\bm{\mathcal{J}}$) and (equivalent) magnetic ($\bm{\mathcal{M}}$) current sources satisfies\footnote{$k_0=\omega\sqrt{\mu_0\epsilon_0}$, $\epsilon_0$, $\mu_0$, $\eta_0=\sqrt{\mu_0/\epsilon_0}$, $\boldsymbol{\bar{\epsilon}}_{r}$, and $\boldsymbol{\bar{\mu}}_{r}$ are the vacuum wave number, vacuum permittivity, vacuum permeability, vacuum plane wave impedance, NBAM relative permittivity tensor, and NBAM relative permeability tensor, respectively. An infinitesimal point/Hertzian dipole current resides at $\bold{r}'=(x',y',z')$, the observation point resides at $\bold{r}=(x,y,z)$, $\Delta \bold{r}=\bold{r}-\bold{r}'=(\Delta x, \Delta y, \Delta z)$, $u(\cdot )$ denotes the Heaviside step function, and $\bold{k}=(k_x,k_y,k_z)$ denotes the wave vector. Furthermore, $\boldsymbol{\bar{\tau}}_{r}=\boldsymbol{\bar{\mu}}_{r}^{-1}$ and $d_0=k_0^2\epsilon_{zz}(\tau_{xy}\tau_{yx}-\tau_{xx}\tau_{yy})$, where $\gamma_{ts}=\bold{\hat{t}}\cdot\boldsymbol{\bar{\gamma}}_{r}\cdot\bold{\hat{s}}$ ($\gamma=\tau,\epsilon$; $t,s=x,y,z$). All derivations are performed for the electric field, but duality in Maxwell's Equations makes immediate the magnetic field solution. Finally, a tilde over variables denotes they are Fourier/wave-number domain quantities.}
\begin{equation}\label{WaveOp} 
\bm{\mathcal{\bar{A}}}(\cdot)=\nabla \times \boldsymbol{\bar{\mu}}^{-1}_{r} \cdot \nabla \times (\cdot) - 
k_0^2 \boldsymbol{\bar{\epsilon}}_{r} 
\cdot (\cdot)\end{equation}
\begin{equation}\bm{\mathcal{\bar{A}}}  (\bm{\mathcal{E}}) = ik_o\eta_o\bm{\mathcal{J}}-\nabla \times 
\boldsymbol{\bar{\mu}}^{-1}_{r} \cdot \bm{\mathcal{M}} \end{equation}
and can be expressed via a 3-D Fourier integral over 
the field's plane wave constituents $\{\bold{\tilde{E}}(\bold{k})\mathrm{e}^{i\bold{k} \cdot \bold{r}}\}$:\footnote{\label{deteq}Adj($\cdot$) and Det($\cdot$) denote the adjugate and determinant of said argument, respectively. Det($\bold{\tilde{\bar{A}}}$)$=d_0(k_z-\tilde{k}_{1z})(k_z-\tilde{k}_{2z})(k_z-\tilde{k}_{3z})(k_z-\tilde{k}_{4z})$, where $\{\tilde{k}_{nz}\}$ are the eigenvalues (i.e., longitudinal [$z$] propagation constants).}
\begin{align}
\bold{\tilde{\bar{A}}}^{-1}&=\mathrm{Adj}\left(\bold{\tilde{\bar{A}}}\right)/\mathrm{Det}\left(\bold{\tilde{\bar{A}}}\right) \\
\bold{\tilde{E}}(\bold{k})&=\bold{\tilde{\bar{A}}}^{-1}\cdot \left[  ik_0\eta_0\bold{\tilde{J}}-\tilde{\nabla} \times \boldsymbol{\bar{\mu}}^{-1}_{r} \cdot \bold{\tilde{M}} \right] \\
\bm{\mathcal{E}}(\bold{r})&= \left(\frac{1}{2\pi}\right)^3\iiint\limits_{-\infty}^{+\infty} \bold{\tilde{E}}(\bold{k}) \, \mathrm{e}^{i\bold{k} \cdot \bold{r}} \, \mathrm{d}k_z \,  \mathrm{d}k_x \, \mathrm{d}k_y \numberthis \label{fte}
\end{align}
where, anticipating planar layering along $z$, the $k_z$ spectral integral is ``analytically" evaluated for every ($k_x,k_y$) doublet manifest in the (typically \emph{numerically}) evaluated outer 2-D Fourier integral. That is to say, by ``analytically" evaluated we mean that the general (symbolic) closed-form solution of the $k_z$ integral for arbitrary $(k_x,k_y)$ doublet, obtained by equivalently viewing the $k_z$ real-axis integral as a contour integral evaluated using Jordan's Lemma and residue calculus, is well-known and can be numerically evaluated at the $(k_x,k_y)$ doublets~\cite{sainath,chew}. In particular, analytical evaluation of the $k_z$ integral yields the ``direct" field $\bm{\mathcal{E}}^d(\bold{r})$~\cite{sainath}:
\begin{multline} 
\bm{\mathcal{E}}^d(\bold{r})=\frac{i}{(2\pi)^{2}} \iint \limits_{-\infty}^{+\infty}\Bigg[u(\Delta z)\sum_{n=1}^2{\tilde{a}_{n}^d\bold{\tilde{e}}_{n}\mathrm{e}^{i\tilde{k}_{nz}\Delta z}} + \\
 u(-\Delta z) \sum_{n=3}^4{\tilde{a}_{n}^d\bold{\tilde{e}}_{n}\mathrm{e}^{i\tilde{k}_{nz}\Delta z}} \Bigg]   \mathrm{e}^{ik_x\Delta x+ik_y\Delta y} \, \mathrm{d}k_x \, \mathrm{d}k_y  \label{Eeqn1} \end{multline}
where $\tilde{a}_{n}^d(k_x,k_y)$ is the (source dependent) direct field amplitude of the $n$th polarization, while $\bold{\tilde{e}}_{n}(k_x,k_y)$ and $\tilde{k}_{nz}(k_x,k_y)$ are (resp.) the electric field eigenvector (i.e., polarization state) and eigenvalue of the $n$th mode ($n=1,2,3,4$)~\cite{sainath}. Modes labeled with $n=1,2$ correspond to up-going polarizations, and similarly for down-going modes ($n=3,4$).\footnote{Please see~\cite{sainath,chew} for other relevant layered-medium expressions.} 

%We show in this note however that reduced expressions, resulting from an optimal choice of spatial basis (namely, the PMB) in which the analytic cancellation of the fictitious double poles of $\bold{\tilde{\bar{A}}}^{-1}$ (and their replacement with simple poles) occurs, %transparently manifests for NBAM.
% Before proceeding to show the NBAM field expressions, we first (i) verify the existence of the plane wave basis in NBAM (including four linearly independent polarization eigenvectors associated with two shared, non-defective eigenvalues) and (ii) show two potential pitfalls %presented by sources radiating within NBAM. 

The problem with the canonical \emph{numerical implementation} of this residue calculus approach lies in its tacit assumption of non-degeneracy (distinctness) in the eigenvalues $\{\tilde{k}_{1z},\tilde{k}_{2z},\tilde{k}_{3z},\tilde{k}_{4z}\}$, which does not hold for NBAM media. As an illustration of the polarization-independent dispersion behavior of NBAM, consider the dispersion relations of a uniaxial-anisotropic medium slab $\{\boldsymbol{\bar{\epsilon}}_{r}=\mathrm{Diag}\left[a,a,b\right],\boldsymbol{\bar{\mu}}_{r}=\mathrm{Diag}\left[c,c,d\right]\}$ ($k_{\rho}^2=k_x^2+k_y^2$)~\cite{felsen,chew}: $\tilde{k}_{1z}=\left[k^2_0ac-(c/d)k_{\rho}^2\right]^{1/2}$, $\tilde{k}_{2z}=\left[k^2_0ac-(a/b)k_{\rho}^2\right]^{1/2}$, $\tilde{k}_{3z}=-\tilde{k}_{1z}$, and $\tilde{k}_{4z}=-\tilde{k}_{2z}$. Setting $\{a=\bar{y}^2c,b=\bar{y}^2d\}$ ($\bar{y}$ is an arbitrary, non-zero multiplicative constant) renders $\tilde{k}_z^+ = \tilde{k}_{1z}=\tilde{k}_{2z}$ and $\tilde{k}_z^- = \tilde{k}_{3z}=\tilde{k}_{4z}$, demonstrating the plane wave propagation direction \emph{dependent}, but polarization \emph{independent}, dispersion characteristics of uniaxial NBAM~\cite{boulanger1}. This conclusion applies also for more general uniaxial NBAM material tensors possessing PMB rotated with respect to the Cartesian basis~\cite{boulanger1}. Similarly, for biaxial NBAM with PMB-expressed tensors $\{\boldsymbol{\bar{\mu}}_{r}^{\mathrm{pmb}}=\mathrm{Diag}\left[a,b,c\right],\boldsymbol{\bar{\epsilon}}_{r}^{\mathrm{pmb}}=\bar{y}^2\boldsymbol{\bar{\mu}}_{r}^{\mathrm{pmb}}\}$, the polarization-independent dispersion relations are:
\begin{align}
\tilde{k}_{1z}^{\mathrm{pmb}}&=\left[(\bar{y}k_0)^2ab-(a/c)k_x^2-(b/c)k_y^2\right]^{1/2}  \numberthis \label{e1}\\
\tilde{k}_{3z}^{\mathrm{pmb}}&=-\left[(\bar{y}k_0)^2ab-(a/c)k_x^2-(b/c)k_y^2\right]^{1/2} \numberthis \label{e2}
\end{align}
with $\tilde{k}_z^+ =\tilde{k}_{2z}^{\mathrm{pmb}}=\tilde{k}_{1z}^{\mathrm{pmb}}$ and $\tilde{k}_z^- =\tilde{k}_{4z}^{\mathrm{pmb}}=\tilde{k}_{3z}^{\mathrm{pmb}}$. 

Now, the two-fold degenerate eigenvalue $\tilde{k}_z^+$ has associated with it two linearly independent field polarizations describing up-going plane waves~\cite{boulanger1}; this holds likewise for the two down-going polarizations with common eigenvalue $\tilde{k}_z^-$. Mathematically speaking, the eigenvalues $\{\tilde{k}_z^+,\tilde{k}_z^-\}$ are each twice-repeating (i.e., algebraic multiplicity of two) but have associated with each of them two linearly independent eigenvectors (i.e., geometric multiplicity of two), making them non-defective and rendering the four NBAM polarization states suitable as a local EM field basis within NBAM layers~\cite{hanson1}. Despite the existence of four linearly independent eigenvectors, it is worthwhile to further exhibit the key results of the systematic analytical treatment of the two fictitious double-poles of $\bold{\tilde{\bar{A}}}^{-1}$ to render numerical PWE-based EM field evaluation robust to the two said sources of ``breakdown"; this treatment is performed in the next section.

Let us first make two preliminary remarks, however. First, assume that the source-containing layer is a biaxial NBAM with $\boldsymbol{\bar{\mu}}_{r}^{\mathrm{pmb}}=\mathrm{Diag}\left[a,b,c\right]$ and $\boldsymbol{\bar{\epsilon}}_{r}^{\mathrm{pmb}}=\bar{y}^2\boldsymbol{\bar{\mu}}_{r}^{\mathrm{pmb}}$. Second, the orthogonal matrix $\bold{\bar{U}}=\begin{bmatrix} \bold{\hat{v}}_1 & \bold{\hat{v}}_2 & \bold{\hat{v}}_3\end{bmatrix}$ transforms vectors between the PMB and Cartesian basis. For example, the relationship between the $n$th PMB eigenmode wave vector $\bold{k}^{\mathrm{pmb}}_{n}=(k_{nx}^{\mathrm{pmb}},k_{ny}^{\mathrm{pmb}},\tilde{k}_{nz}^{\mathrm{pmb}})$ and the (assumed available\footnote{The Cartesian basis wave vectors and polarization eigenvectors are assumed available (e.g., via the state matrix method~\cite{chew}). Indeed, recall that the operations discussed herein are performed within the backdrop of numerical 2-D Fourier integral evaluations~\cite{sainath}.}) $n$th Cartesian-basis wave vector $\bold{k}_{n}=(k_x,k_y,\tilde{k}_{nz})$ writes as $\bold{k}^{\mathrm{pmb}}_{n}=\bold{\bar{U}}^{-1}\cdot\bold{k}_{n}$.

\section{\label{form}Direct Electric Field Radiated within NBAM}

The (Cartesian basis) Fourier domain representation of the electric field, radiated in a homogeneous NBAM, writes as $\bold{\tilde{E}}=-\bold{\tilde{\bar{A}}}^{-1} \cdot \bold{\tilde{\nabla}} \times \boldsymbol{\bar{\mu}}^{-1}_{r}\cdot \bold{\tilde{M}}$ for a (equivalent) magnetic current source or $\bold{\tilde{E}}= ik_0\eta_0 \bold{\tilde{\bar{A}}}^{-1} \cdot\bold{\tilde{J}}$ for an electric current source. These two equations, moreover, hold equally when re-represented in the NBAM's PMB (i.e., adding ``pmb" superscript to all quantities), which is what we will employ. Indeed, the components $\{A_{mw}\}$ ($m,w=1,2,3$) of $\bold{\tilde{\bar{A}}}^{-1,\mathrm{pmb}}(\cdot)$ write as ($A_{mw}= A_{wm}$, and $\bold{\bar{k}}=\bold{k}^{\mathrm{pmb}}/k_0$):
\begin{align}
\tilde{B}&= -c\bar{y}^2k_0^2\left(\bar{k}_z^2-\left[ab\bar{y}^2-(a/c)\bar{k}_x^2-(b/c)\bar{k}_y^2\right]\right) \numberthis \label{exp1}\\
A_{11}&= \left(\bar{k}_x^2-bc\bar{y}^2\right)/\tilde{B}, \ A_{12}= \bar{k}_x\bar{k}_y/\tilde{B} \\ 
A_{13}&= \bar{k}_x\bar{k}_z/\tilde{B}, \ A_{22}= \left(\bar{k}_y^2-ac\bar{y}^2\right)/\tilde{B} \\ 
A_{23}&= \bar{k}_y\bar{k}_z/\tilde{B}, \ A_{33}=\left(\bar{k}_z^2-ab\bar{y}^2\right)/\tilde{B} \numberthis \label{exp2}
\end{align}
while the components of $-\bold{\tilde{\bar{A}}}^{-1,\mathrm{pmb}}\cdot\bold{\tilde{\nabla}}^{\mathrm{pmb}}\times\boldsymbol{\bar{\mu}}^{-1,\mathrm{pmb}}_{r}(\cdot)$ $\{ \dot{A}_{mw} \}$ write as ($\dot{A}_{mw}=-\dot{A}_{wm}$):
\begin{align}
\tilde{B}'&= \tilde{B}/(\bar{y}^2), \ \dot{A}_{12}=-ick_z/\tilde{B}' \numberthis \label{exp3a} \\
\dot{A}_{13}&=ibk_y/\tilde{B}', \ \dot{A}_{23}=-iak_x/\tilde{B}'  \numberthis \label{exp3}
\end{align}
The expressions within Eqns. \eqref{exp1}-\eqref{exp2} describe the electric field from an electric current source while the expressions within Eqns. \eqref{exp3a}-\eqref{exp3} describe the electric field from an (equivalent) magnetic current source. Duality in Maxwell's Equations makes immediate the magnetic field results. 

%\footnote{Strictly speaking, the up-going (down-going) modes are realized only for $z>z'$ ($z<z'$), while the $z=z'$ case results in an improperly defined $k_z$ contour integral.\cite{chew} However, since we assume the source does not lie exactly at a material interface, %physical arguments lead to the ($z=z'$)-plane field distribution arising as the (common) limiting case of the field distributions on planes $z= (z'+\delta)$ and $z= (z'-\delta)$ as $\delta \to 0^+$. Therefore, subject to using properly defined numerical integration paths in the %$k_x$ and $k_y$ planes,\cite{sainath3} one can compute the ($z=z'$)-plane direct fields through extracting either the up-going or down-going direct fields and evaluating them at $z=z'$.}
%\mathrm{e}^{ik_x\Delta x+ik_y\Delta y+i\tilde{k}^+\Delta z}

Next the PMB electric field $\bold{E}^{\mathrm{pmb}}(k_x,k_y;z,z')$, after re-expressing Eqns. \eqref{exp1}-\eqref{exp3} in terms of $\{k_x,k_y,k_z\}$ to identify the $k_z$ (rather than $k_z^{\mathrm{pmb}}$) eigenvalues $\{\tilde{k}_{nz}\}$ (using the relation $\bold{k}=\bold{\bar{U}}\cdot \bold{k}^{\mathrm{pmb}}$) as well as ``analytically" performing the $k_z$ contour integral, can be decomposed into a linear combination of the degenerate up-going modes $\{\bold{\tilde{e}}_1^{\mathrm{pmb}},\bold{\tilde{e}}_2^{\mathrm{pmb}}\}$ (for $z > z'$) or down-going modes $\{\bold{\tilde{e}}_3^{\mathrm{pmb}},\bold{\tilde{e}}_4^{\mathrm{pmb}}\}$ (for $z < z'$).\footnote{When $z=z'$, assuming the source does not lie exactly at a planar material interface, one can write the direct fields as a linear combination of either the up-going \emph{or} down-going modes since both combinations lead to identical field results (save at $\bold{r}'$) on the plane $z=z'$~\cite{sainath,chew}.} For an electric source, we have for $\bold{\tilde{e}}^{\pm,\mathrm{pmb}}$:
\begin{equation}
\pm 2\pi i\left[ik_0\eta_0\left(k_z-\tilde{k}_z^{\pm}\right)\bold{\tilde{\bar{A}}}^{-1,\mathrm{pmb}}\cdot \bold{\tilde{J}}^{\mathrm{pmb}}\right]\Bigg|_{k_z=\tilde{k}_z^{\pm}} \numberthis \label{e1a} 
\end{equation}
and similarly for a (equivalent) magnetic source upon replacing $ik_0\eta_0\bold{\tilde{\bar{A}}}^{-1,\mathrm{pmb}}\cdot \bold{\tilde{J}}^{\mathrm{pmb}}$ with $-\bold{\tilde{\bar{A}}}^{-1,\mathrm{pmb}}\cdot\bold{\tilde{\nabla}}^{\mathrm{pmb}} \times \boldsymbol{\bar{\mu}}^{-1,\mathrm{pmb}}_{r} \cdot \bold{\tilde{M}}^{\mathrm{pmb}}$ in Eqn. \eqref{e1a}. Next, the degenerate PMB modal electric fields are re-expressed in the Cartesian basis ($\bold{\tilde{e}}^{\pm}=\bold{\bar{U}}\cdot\bold{\tilde{e}}^{\pm,\mathrm{pmb}}$) from which the Cartesian-basis direct field modal amplitudes $\{\tilde{a}_{1}^{d},\tilde{a}_{2}^{d},\tilde{a}_{3}^{d},\tilde{a}_{4}^{d}\}$ can be robustly extracted using the polarization decomposition method proposed previously for sources radiating within isotropic layers~\cite{sainath}:
\begin{equation}
\begin{bmatrix}
\tilde{a}_{1}^{d} \\
\tilde{a}_{2}^{d}
\end{bmatrix}=\begin{bmatrix}
\tilde{e}_{x1} & \tilde{e}_{x2} \\
\tilde{e}_{y1} & \tilde{e}_{y2}
\end{bmatrix}^{-1}
\begin{bmatrix}
\tilde{e}^{+}_x \\
\tilde{e}^{+}_y
\end{bmatrix},\ 
\begin{bmatrix}
\tilde{a}_{3}^{d} \\
\tilde{a}_{4}^{d}
\end{bmatrix}=
\begin{bmatrix}
\tilde{e}_{x3} & \tilde{e}_{x4} \\
\tilde{e}_{y3} & \tilde{e}_{y4}
\end{bmatrix}^{-1}
\begin{bmatrix}
\tilde{e}^{-}_x \\
\tilde{e}^{-}_y
\end{bmatrix}
\label{decomp}
\end{equation}
where $\{\tilde{e}_{xn},\tilde{e}_{yn},\tilde{e}_{zn}\}$ are the $x$, $y$, and $z$ components of the (cartesian basis) NBAM's $n$th electric field eigenvector $\bold{\tilde{e}}_n$. Moreover, if the above-inverted matrices are suspected (with respect to, say, the euclidean matrix norm measure) of being ill-conditioned, one can always utilize instead say the $y$ and $z$, or alternatively the $x$ and $z$, components of the field eigenvectors~\cite{sainath}. Indeed, this decomposition procedure is well-defined due to the non-defective nature of the eigenvalues, and hence linear independence between the four NBAM field eigenvectors $\{\bold{\tilde{e}}_{n}\}$~\cite{boulanger1}.

\section{\label{res}Results}
\begin{figure}[H]
\centering
\includegraphics[width=2in]{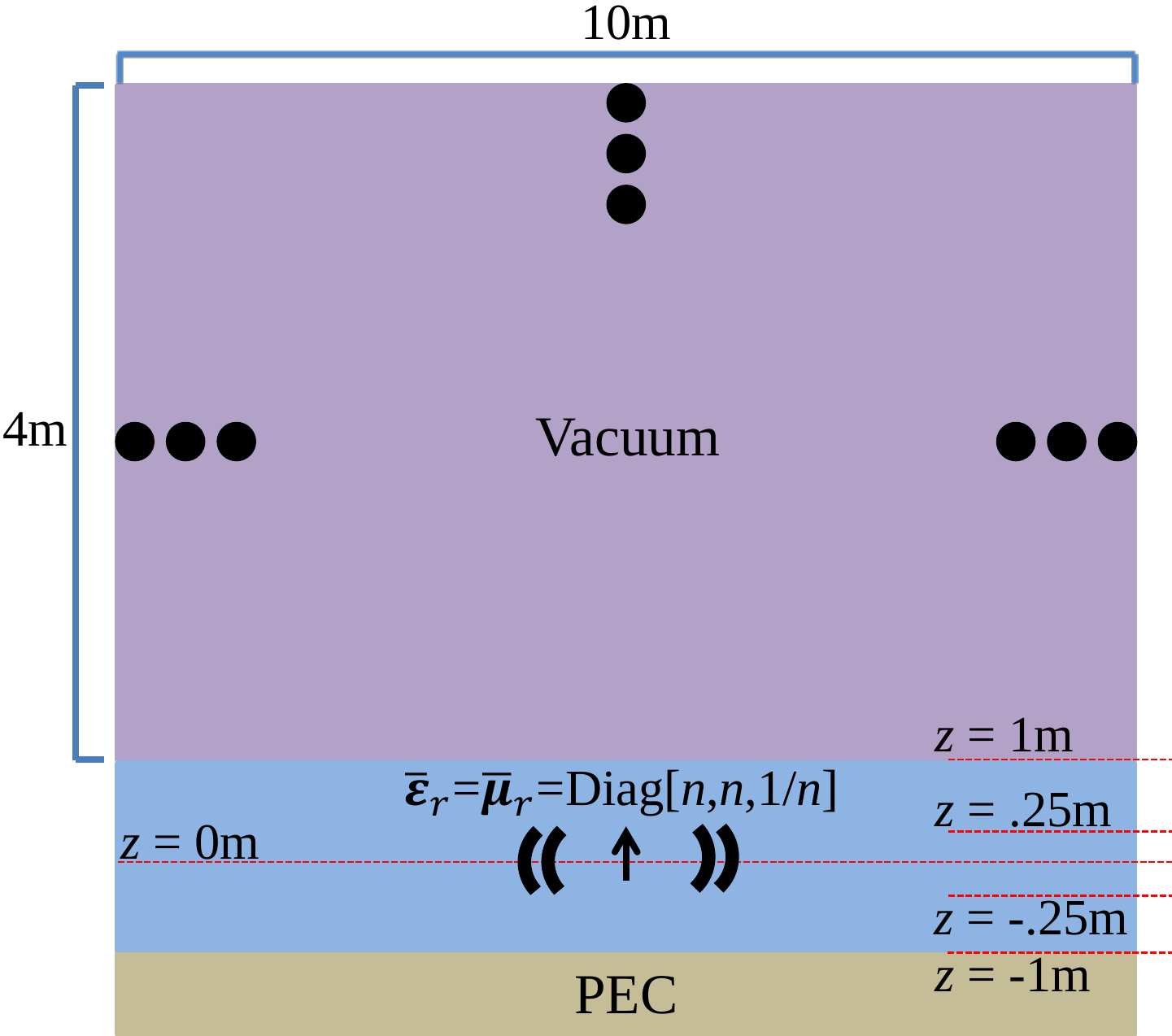}
\caption{\label{geo1}\small Vertically-oriented Hertzian dipole current source within a three-layer NBAM. The purple (air) and blue (NBAM) regions form the plane on which the fields are observed in Fig. \ref{f2}. The parameter $n$ equals ten, five, and two within the regions $-1 < z < -1/4$ [m], $-1/4 < z < 1/4$ [m], and $1/4 < z < 1$ [m], respectively.}
\end{figure}

Now we exhibit some illustrative results demonstrating the developed algorithm's performance. We investigate both the electric field $\mathcal{E}_z$ radiated by a vertical (i.e., $z$-directed) Hertzian electric current dipole (VED), as well as the magnetic field $\mathcal{H}_z$ radiated by a $z$-directed Hertzian (equivalent) magnetic current dipole (VMD); both sources radiate at $f=2$MHz. In both scenarios, the source resides at depth $z'=0$m within a three-layer NBAM, occupying the region $-1 \leq z \leq 1$ [m], of material properties $\boldsymbol{\bar{\epsilon}}_r=\boldsymbol{\bar{\mu}}_r=\mathrm{Diag}[10,10,1/10]$, $\mathrm{Diag}[5,5,1/5]$, and $\mathrm{Diag}[2,2,1/2]$ within the regions $-1 < z < -1/4$ [m], $-1/4 < z < 1/4$ [m], and $1/4 < z < 1$ [m] (resp.); see Fig. \ref{geo1}. The top layer ($z \geq 1$m) is vacuum ($\epsilon_{r1}=\mu_{r1}=1$) while the bottom layer ($z \leq -1$m) is a perfect electric conductor (PEC); note that this layered-medium configuration was specifically chosen to facilitate comparison with closed-form solutions through invocation of T.O. and EM Image theory~\cite{sainath4}. Indeed the EM field solution within $z\geq -1$m, for our five-layered configuration involving a VED source, can be shown identical to the closed-form field result of two VED's (located at depths $z=-1.75$m and $z=-19.25$m) of identical orientation to the original VED and radiating in homogeneous, unbounded vacuum. Note that within the NBAM, an added step to compute the closed-form result must be taken, appropriately mapping the observation points within the NBAM to vacuum observation points by viewing a $d$-meter thick NBAM layer $\boldsymbol{\bar{\epsilon}}_r=\boldsymbol{\bar{\mu}}_r=\mathrm{Diag}[n,n,1/n]$ as equivalent to a $nd$-meter thick vacuum layer. Similarly, the VMD problem can be shown identical to two VMD's (located at depths $z=-1.75$m and $z=-19.25$m) radiating in homogeneous, unbounded vacuum; in this scenario however, image theory prescribes that the $z=-1.75$m VMD possess identical orientation to the original VMD, but that the $z=-19.25$m VMD possess \emph{opposite} orientation.\footnote{The amplitudes of the VED and VMD (i.e., lying within the central NBAM layer) must be scaled by a factor of 1/5 (relative to the vacuum sources) to facilitate field comparisons. Moreover the normal field components $\{\mathcal{E}_z,\mathcal{H}_z\}$, within the NBAM layer with properties $\boldsymbol{\bar{\epsilon}}_r=\boldsymbol{\bar{\mu}}_r=\mathrm{Diag}[n,n,1/n]$, are also scaled (artificially, for both visual display and error computation purposes) by $1/n$ to account for their discontinuity across material interfaces.}

\begin{figure}[h]
\centering
\subfloat[\label{geom3z}]{\includegraphics[width=1.75in,height=1.35in]{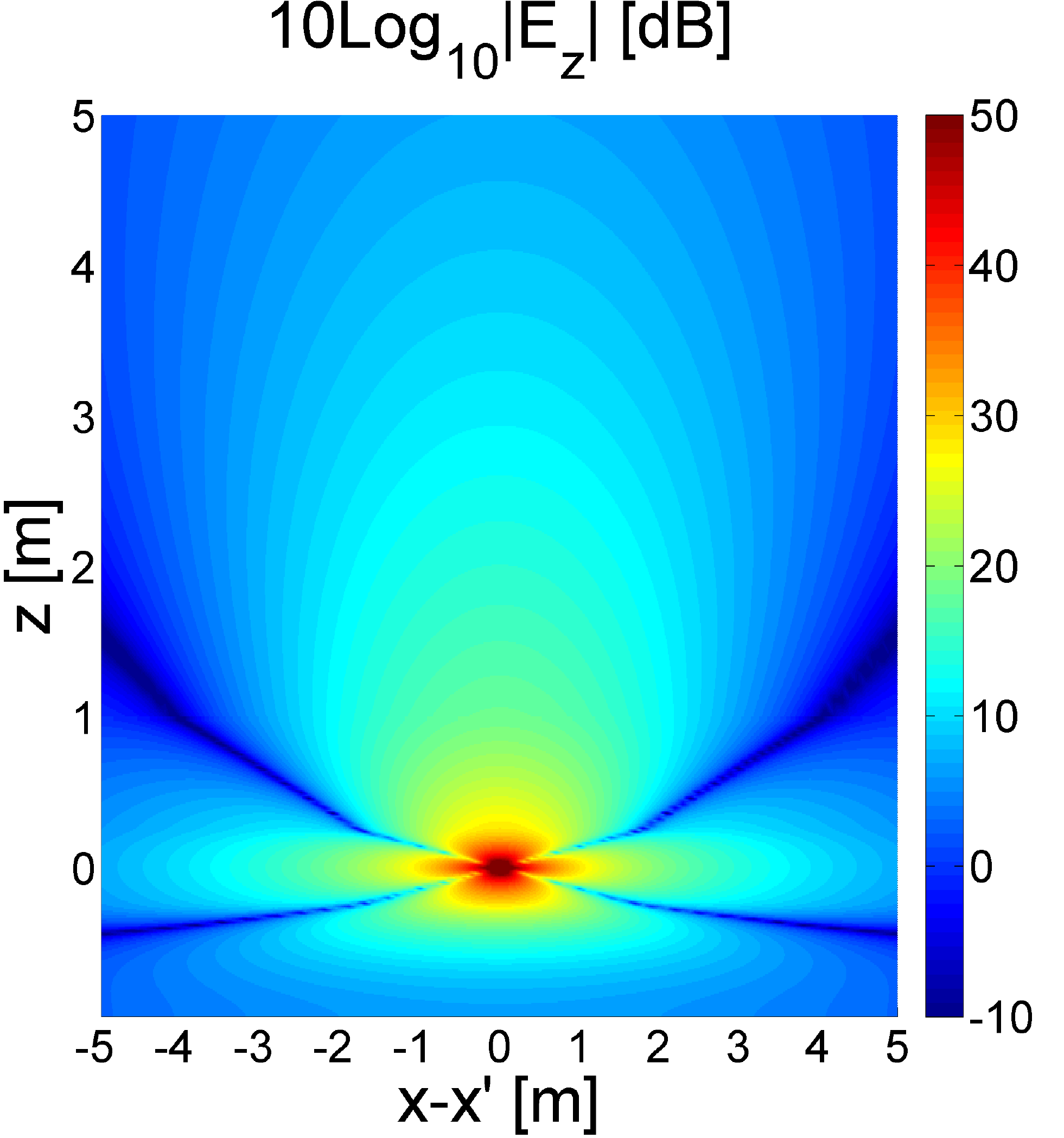}}
\subfloat[\label{geom3a}]{\includegraphics[width=1.75in,height=1.35in]{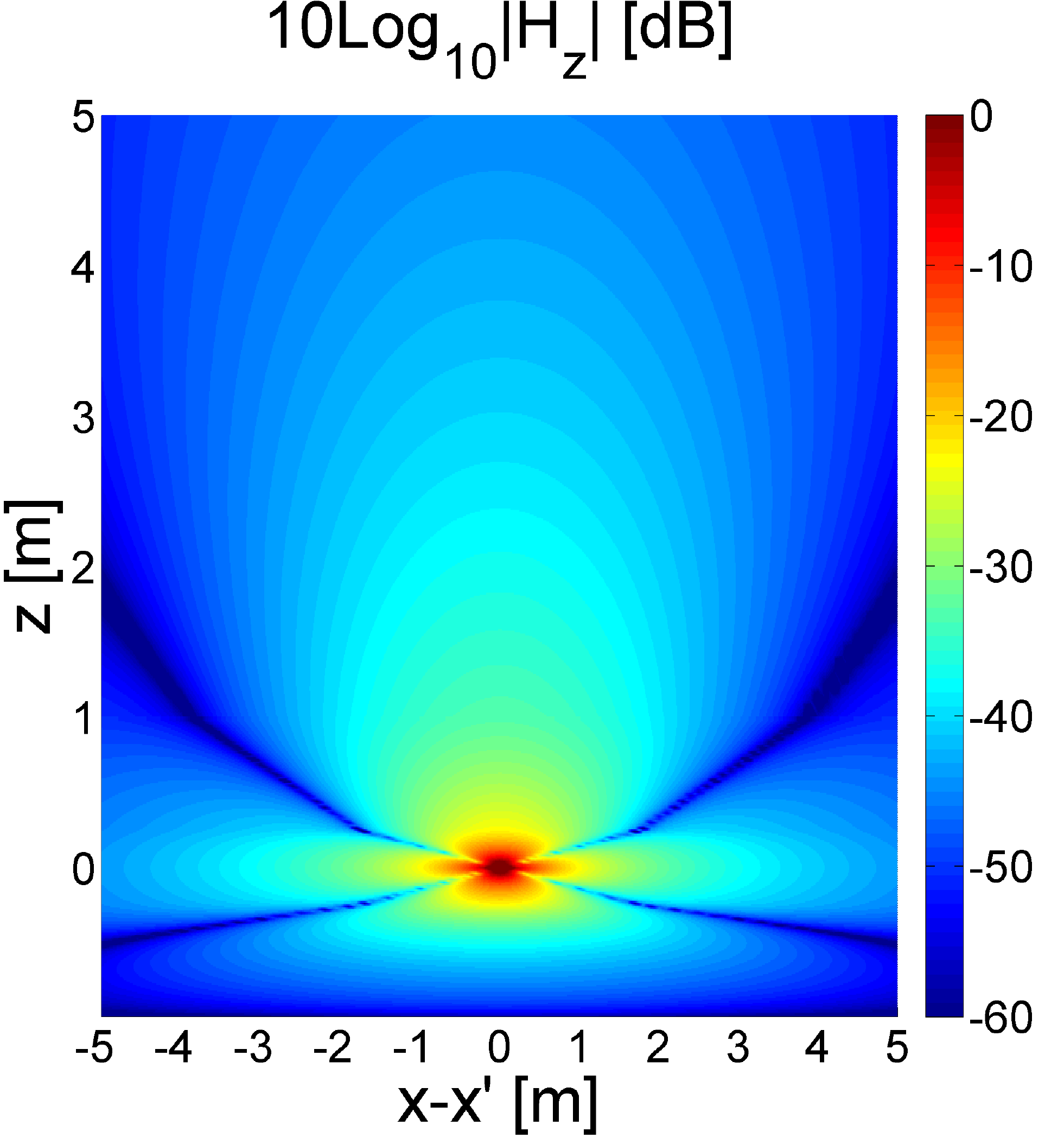}}

\subfloat[\label{geom3c}]{\includegraphics[width=1.75in,height=1.35in]{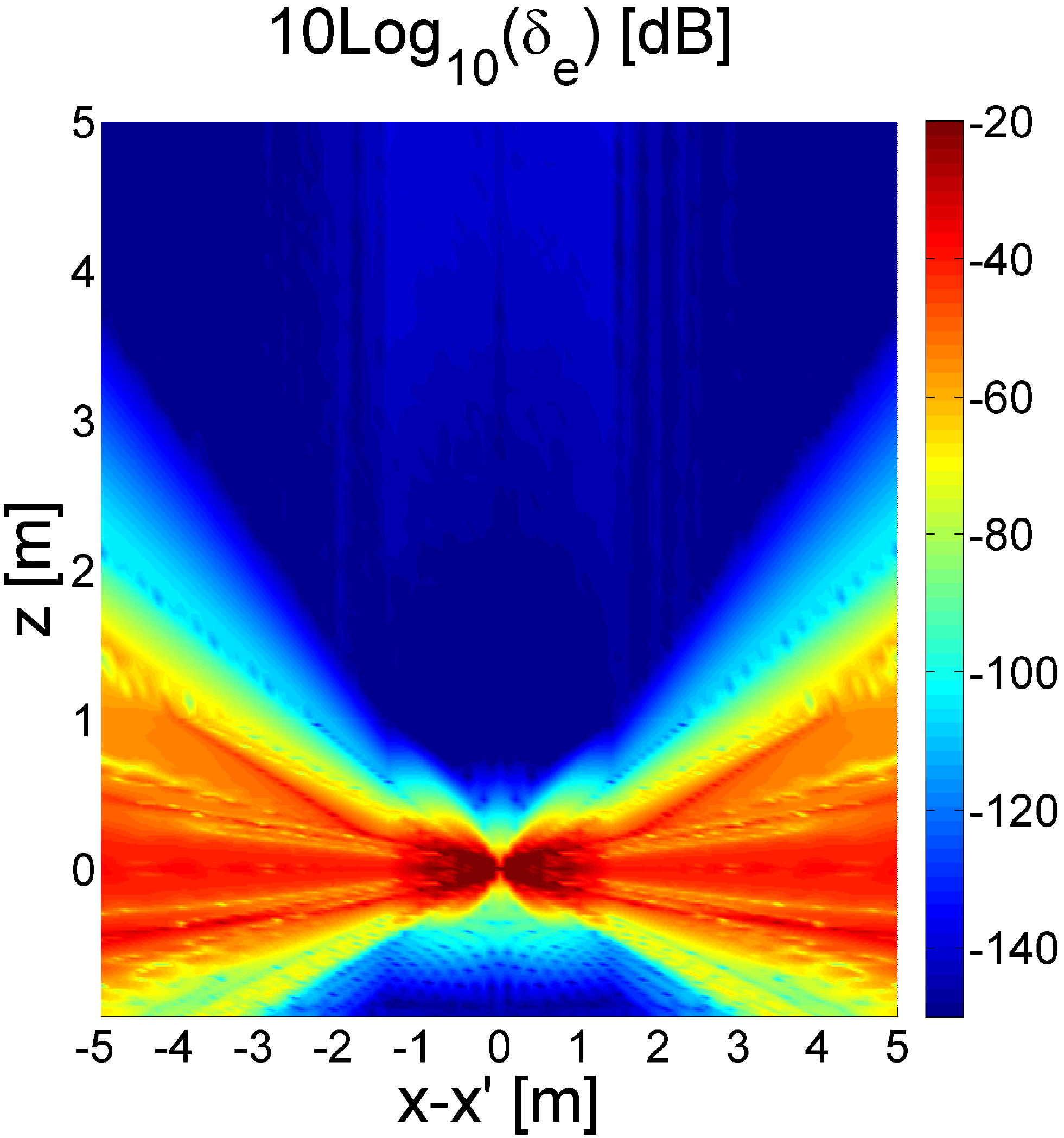}}
\subfloat[\label{geom3e}]{\includegraphics[width=1.75in,height=1.35in]{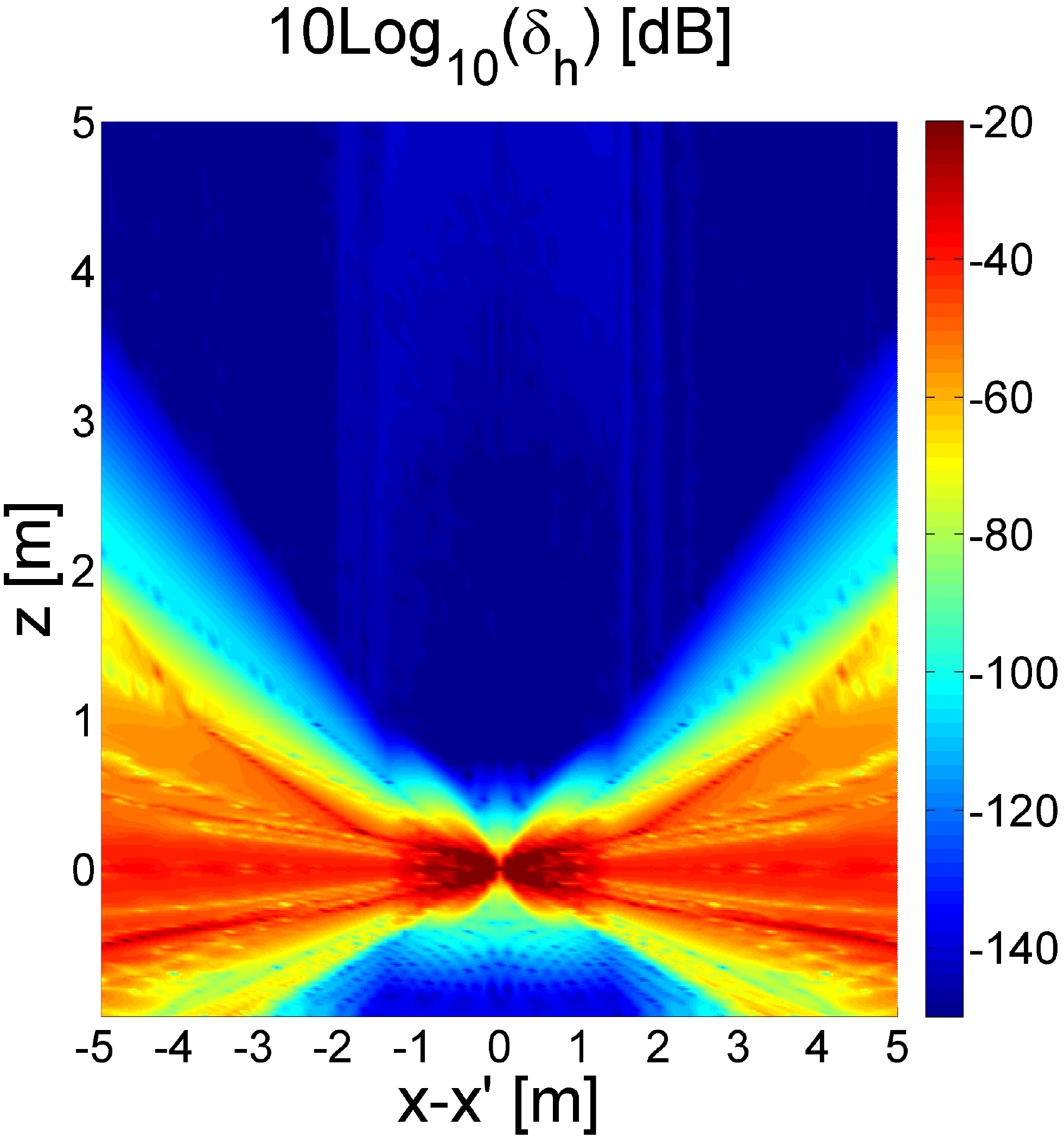}}
\caption{\label{f2}\small (a) $\mathcal{E}_z$ radiated by a VED. (b) $\mathcal{H}_z$ radiated by a VMD. (c) Relative error: $\mathcal{E}_z$. (d) Relative error: $\mathcal{H}_z$.}
\label{geom3}
\end{figure}

Observing Figs. \ref{geom3c}-\ref{geom3e}, we note the relative errors in both the electric field ($\delta_e$) and magnetic field ($\delta_h$) are very low,\footnote{Let $E_c$ be the computed electric field, and let $E_v$ be the closed-form reference solution. Then $\delta_e=|E_c-E_v|/|E_v|$ (likewise for $\delta_h$).} approaching in most of the observation plane near the limits of floating point double precision-related numerical noise (approximately -150 in [dB] scale); for reference, Figs. \ref{geom3z}-\ref{geom3a} are the computed field distributions themselves from our algorithm. This is consistent with our having set an adaptive relative integration error tolerance of $~1.2 \times 10^{-14}$. We do observe however that the error noticeably increases (for fixed observer/source radial separation) as the observation angle tends closer to ``horizon" (i.e., source depth $z'$ and observer depth $z$ coinciding). The error variation trend versus angle has been observed before~\cite{sainath4} even when the source resided in non-NBAM media, and hence the increased error versus observation angle is not likely due to instabilities in the presented NBAM-robust algorithm. We conjecture rather that the increasing error (versus observation angle) arises due to commensurately increasing numerical cancellation\footnote{Namely, cancellation from radiation field contributions arising from numerical integration along contour sub-sections symmetrically located about the imaginary $k_x$ and $k_y$ axes. By contrast, our algorithm robustly ensures (irrespective of observation angle) that the \emph{evanescent} field contribution introduces little numerical cancellation-induced error and rapid convergence~\cite{sainath3}.} that can only be partially offset by a (computer resource limited) \emph{finite} extent of $hp$ integration refinement performed using \emph{finite} precision arithmetic. This numerical cancellation, we remark, is well known to be predominantly induced by integrand oscillation, which worsens as the observation angle tends to horizon~\cite{sainath4,sainath}. One remedy is to use a constant-phase path~\cite{lambot1}, but a robust remedy for 2-D integrals (needed for generally anisotropic media) remains an open question. Moreover, this path would change as one varies the outer integration variable. Finally we emphasize that given the design of our \emph{particular} implementation, which always first computes the direct electric field and \emph{then} (if need be) computes the magnetic field using ancillary relations~\cite{chew}[Ch. 2], we have in fact tested the soundness of both Eqns. \eqref{exp1}-\eqref{exp2} (VED scenario) and Eqns. \eqref{exp3a}-\eqref{exp3} (VMD scenario). 

\section{\label{conc}Conclusion}
We addressed a fundamental origination of breakdown in the spectral-domain-based (PWE) evaluation of EM fields radiated by sources embedded within NBAM planar slabs, leading to a robust formulation that can accurately compute EM fields despite the modal degeneracy, induced by said NBAM, that would ordinarily lead to numerical instabilities and/or corruption of the functional form of the plane wave eigenfunctions. Indeed this instability arises due to eigenvalues that, while non-defective, have an algebraic multiplicity equal to two rather than one. The remedy is to apply a proper (analytical) ``pre-treatment" of the spectral-domain tensor operators prior to polarization amplitude extraction, resulting in robust analysis of EM fields in arbitrary anisotropic planar-layered media. Results validated the high accuracy of numerical computations based on this analytical pre-treatment.
\bibliography{reflist}
\bibliographystyle{IEEEtran}
\end{document}